\documentclass[fleqn,12pt,twoside]{article}
\usepackage{espcrc1}

\usepackage{epsfig}

\title{Diffractive Dijet Production and Nuclear Shadowing 
	in $pA$ Interactions}

\author{S.E. Vance and D. Kharzeev
	\address{Physics Department, Brookhaven National Laboratory, 
	         Upton, NY 11973}
      	\thanks{We would like to thank A. Capella, B. Kopeliovich 
and K. Goulianos for their comments. 
This manuscript was authored under Contract No. DE-AC02-98CH10886 
with the U. S. Department of Energy.}}

\begin{document}
\maketitle

\begin{abstract}
We study the implications of non-universality observed recently
in $ep$ and $\bar{p}p$ diffraction for nuclear shadowing and diffractive 
dijet production in $pA$ collisions.
\end{abstract}

\section{Nuclear Shadowing}
Recently, measurements of diffractive jet rates in $\bar{p}p$ interactions
at the Tevatron were observed to be significantly lower than the 
theoretical expectations (for a review see \cite{abramowicz00}).   
The expectations were obtained using parameterizations of 
diffractive parton distribution functions extracted 
from $ep$ interactions at HERA.  
The discrepancy between the data and the calculations revealed the breakdown 
of factorization or the onset of absorptive corrections in 
$\bar{p}p$ diffractive events\cite{collins98}.   

The physical reason for the breakdown of factorization is likely to be  
related to the different color structure of the probes in $\bar{p}p$ and $ep$ 
reactions. In $ep$, the compact, color singlet $\bar{q}q$ pair 
produced by the virtual photon has a smaller probability 
of interacting with the external color fields and therefore has a larger 
rapidity gap survival probability. In $\bar{p}p$, the color octet 
gluon at comparable virtualities still interacts strongly with the
external color fields, reducing its rapidity gap 
survival probability.   
The implications of the breakdown of factorization are now 
studied in nuclear shadowing and dijet production.

Nuclear shadowing is defined as the ratio of the $pA$ to 
the $pp$ total cross section, $R = \sigma_{pA}/A \sigma_{pp}$.  
The total cross section for $pA$ interactions can be 
written as the sum of 
two terms
\begin{equation}
\sigma_{pA} \simeq A \sigma_{pN} - \delta \sigma_{pA},
\end{equation}
where the second term, representing elastic and inelastic
\cite{gribov69} shadowing, can be written as\cite{gribov69,karmanov73} 
\begin{eqnarray}
\delta \sigma_{p A} &\simeq& - 8 \pi \int d^2b \int_{-\infty}^{\infty} 
dz_1 \int_{z_1}^{\infty} dz_2 \;\; \rho_A(\vec{b},z_1) \rho_A(\vec{b},z_2)  
\label{deltaSigma_pA} \\ \nonumber 
&& \times \int_{M_0^2}^{0.1s} dM^2 \cos \left((z_1 - z_2)/ \lambda  
\right ) \left. \frac{ d^2 \sigma^{sd}_{pN}} { dM^2 dt}
\right|_{t \approx 0} \exp \left[ -\frac{\sigma_{XN}}{2} 
\int_{z_1}^{z_2} dz \rho_A(\vec{b},z) \right], 
\end{eqnarray} 
where $d^2 \sigma^{sd}_{pN}/dM^2 dt$ is the single diffractive
cross section, $\lambda = m_p (M^2 - m_p^2) / s$ is the coherence
length of the resonance state X, and 
$\rho_A(\vec{r})$ is the nuclear density 
($\rho_A$ is normalized to $A$ and is taken to be Gaussian 
in all calculations presented here).  In Eq. \ref{deltaSigma_pA}, $\delta 
\sigma$ depends upon the $pp$ diffractive cross section.  
Unlike $\gamma^*A$ interactions, 
$R$ can not be directly associated with the shadowing of the 
structure functions (for a review see \cite{pillerRev}).

In Regge theory, the dominant contribution to 
the single diffractive cross section at high energy is 
the triple Pomeron term.  The single diffractive cross section 
can be expressed as 
\begin{equation}
\frac{d^2 \sigma^{sd}}{d(M^2/s) dt} = f_{P/p}(M^2/s,t)\sigma_T^{Pp}(M^2/s_0), 
\end{equation}
where 
$f_{P/p}(M^2/s,t) = \frac{\beta^2_{Ppp}(t)}{16 \pi}  
\left ( \frac{M^2}{s} \right )^{1 - 2 \alpha_P(t)}$ and 
$ \sigma_T^{Pp}(M^2/s_0) = 
\beta_{Ppp}(0) g(t) \left(\frac{M^2}{s_0}\right)^{\alpha_p(0) - 1}$.
In these expressions, $\beta_{Ppp}(t)$ represents the coupling of the Pomeron 
to the proton, $s_0 = 1 \; \mbox{GeV}$, and the Pomeron intercept
$\alpha_P(t) = \alpha(0) + \alpha' t = 1 + \epsilon + 0.25t$,
where $\epsilon \sim 0.08-0.10$.   While this form is able to describe the  
diffractive $ep$ cross section, it fails to describe 
the recently observed energy dependence of the 
diffractive $\bar{p}p$ cross section. 

A reasonable parameterization of the diffractive $\bar{p}p$ cross section
is obtained by normalizing $f_{P/p}(M^2/s,t)$ 
according to the following scheme\cite{goulianos99}, 
\begin{equation} 
f_N(M^2/s,t) = \left \{ \begin{array}{ll}
f_{P/p}(M^2/s,t) & \mbox{if $N(s) < 1$} \\ 
\frac{1}{N(s)} f_{P/p}(M^2/s,t) & \mbox{ if $N(s) > 1$}
\end{array} \right. \label{flux} 
\end{equation}
where 
$N(s) = \int^{0.1}_{1.5/s} d(M^2/s) \int_{- \infty}^{0} dt 
f_{P/p}(M^2/s,t).$ 
In this scheme, the cross section becomes constant at high energies,
$\lim_{s \rightarrow \infty} \sigma^N_{sd}(s) = constant$. 

\begin{figure}[thb]
\epsfxsize=2.0in   
\epsfbox{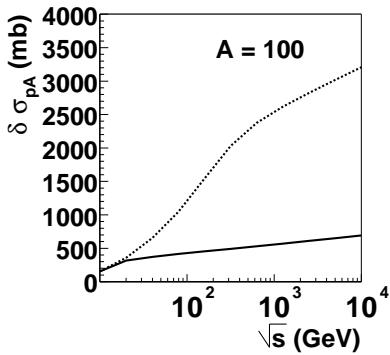}
\caption{ Calculations of the correction term, $\delta \sigma_{pA}$, 
for $A=100$ with (solid line) and without (dashed line) the  
normalization scheme of Eq. \protect\ref{flux}.}
\label{delta_sd}
\end{figure}

The effect of using this parameterization on the correction term in 
Eq. \ref{deltaSigma_pA} is shown in Figure \ref{delta_sd}.  
Here, the solid line is a calculation of $\delta \sigma$ for $A=100$ with 
the parameterization, while the dashed line is the 
n\"aive triple Pomeron contribution.   
The normalization scheme needed to reproduce $\bar{p}p$ data 
results in a dramatic reduction of the shadowing in $pA$.

\section{Diffractive dijet production} 
A simple formula for calculating diffractive dijet production 
will be utilized in this section.  To begin, the hadronic state
$|h_l \rangle$ is expanded in terms of its diffractive eigenstates
$| \alpha \rangle$ \cite{grassberger77,miettinen78}, 
\begin{equation}
|h_l \rangle = \sum_{\alpha} C_{l \alpha} | \alpha \rangle 
\end{equation}
         
The diffractive cross section can be expressed 
as the sum over the cross sections for all diffractive final states
minus the elastic cross section, 
\begin{eqnarray}
\sigma^{diff}_{p A} &=& \sum_l \sigma(p A \rightarrow h_l A) 
- \sigma^{el}_{p A} \\ 
&=& \int d^2 b \left [  \sum_l \left | C_{l \alpha} \right |^2 
e^{-\sigma_{\alpha}T(\vec{b})}  
- \left ( \sum_l \left | C_{l \alpha} \right|^2 e^{-\sigma_{\alpha}T(\vec{b})} 
\right )^2 \right ], \label{diff_pA}
\end{eqnarray}
where $ T(\vec{b}) = \int_{-\infty}^{\infty} dz \rho(\vec{b},z)$ is the nuclear
thickness function.

Assuming that $|C_{l\alpha}|^2 = 1/2$, the cross sections are related 
according to $\sigma_{\alpha} = \sigma_{tot} \pm \Delta,$ where 
$\Delta^2 = 16\pi \int dM^2 \frac{d\sigma}{dM^2 dt},$  
and the terms in Eq. \ref{diff_pA} can then be written as 
\begin{equation}
\sum_{\alpha} |C_{l\alpha}|^2 e^{\frac{1}{2} \sigma_\alpha T(\vec{b})} 
= \frac{1}{2} e^{ \frac{1}{2} \sigma_{tot} T(\vec{b}) } 
\left ( e^{- \frac{1}{2}\Delta \tilde{T}^2(\vec{b},M^2)} + 
e^{\frac{1}{2}\Delta \tilde{T}^2(\vec{b},M^2)} \right ), \label{sum}
\end{equation}
where $\tilde{T}(\vec{b},M^2) = \int_{-\infty}^{\infty} dz \rho(\vec{b},z) 
e^{i z/ \lambda(s,M^2)}$. 

\begin{figure}[htb]
\epsfxsize=3.0in   
\epsfbox{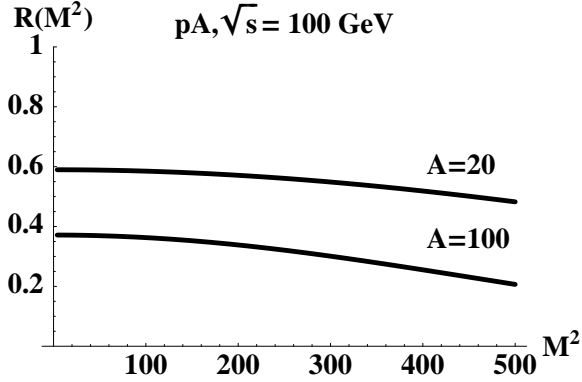}
\caption{ Calculations of the ratio, $R(M^2)$, of the diffractive cross 
sections of $pA$ to $pp$ are shown 
for $A = 20$ and $A = 100$ at $\sqrt{s} = 100 \; \mbox{GeV}$.}
\label{ratio_pA}
\end{figure}

Substituting Eq. \ref{sum} into Eq. \ref{diff_pA},
the diffractive cross section can be expressed as
\begin{equation}
\sigma_{p A}^{diff} 
= 4 \pi \int d^2 b \;\int dM^2 \; \tilde{T}^2(\vec{b},M^2) \; 
e^{ - \sigma^{tot}_{pp } T(\vec{b})} \;
\left. \frac{d\sigma_{pp}}{dM^2 dt} \right |_{t=0}, \label{sig_pA_dif}
\end{equation}
and the ratio of the $pA$ to $pp$ differential diffractive cross section
is 
\begin{equation}
R(M^2) = \left. \frac{d\sigma_{pA}^{diff}}{dM^2} \right / 
 A \frac{d\sigma_{pp}^{diff}}{dM^2}
= \frac{ 4 \pi }{A} \int d^2 b \; \tilde{T}^2(\vec{b},M^2) \; 
e^{ - \sigma_{pp}^{tot} T(\vec{b})}. \label{ratio}
\end{equation}
At large $M^2$, Eq. \ref{ratio} represents the ratio of the diffractive dijet
production in $pA$ to $pp$.  In Figure \ref{ratio_pA}, $R(M^2)$ is plotted 
for $A=20$ (higher curve) and for $A=100$ (lower curve) at 
$\sqrt{s} = 100 \; \mbox{GeV}$.  As A increases, fewer diffractive dijets 
are produced relative to $pp$.

\begin{figure}[htb]
\epsfxsize=3.0in   
\epsfbox{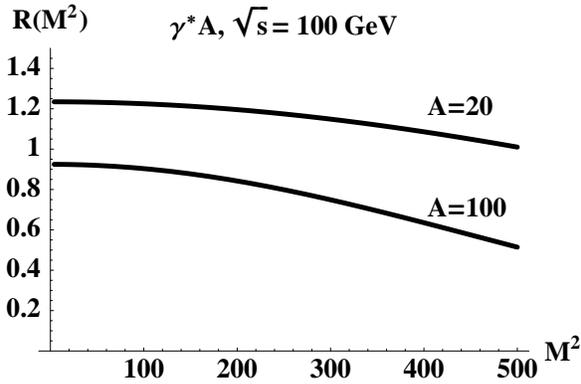}
\caption{ Calculations of the ratio, $R(M^2)$, of the diffractive cross 
sections of $\gamma^* A$ to $\gamma^* p$ are shown for $A = 20$ and $A = 100$
at $\sqrt{s} = 100 \; \mbox{GeV}$.}
\label{ratio_eA}
\end{figure}

A similar formula applies for $\gamma^*A$ diffractive dijet production
(changing $\frac{d\sigma_{pp}^{diff}}{dM^2} \rightarrow 
\frac{d\sigma_{\gamma^* p}^{diff}}{dM^2}$ and 
$\sigma_{pp}^{tot} \rightarrow \sigma_{\gamma^* p}^{tot}$).
The ratio for $\gamma^*A$ is shown in Figure \ref{ratio_eA}.  
The ratio of dijets with the $\gamma^*$ probe is larger than in 
$pA$, since $\sigma_{pp}^{tot} > \sigma_{q\bar{q} p}^{tot}$.
    
Measurements of diffractive dijet production  
and nuclear shadowing in $pA$ interactions at RHIC energies
are needed to better understand the nature of diffraction and the quark-gluon 
wave function of the nucleus.


\begin{thebibliography}{99}
\bibitem{abramowicz00}
H. Abramowicz, Int. J. Mod. Phys. A15S1 (2000) 495 [hep-ph/0001054]. 

\bibitem{collins98}
J. C. Collins, Phys. Rev. D57 (1998) 3051 [hep-ph/9709499]; 
erratum Phys. Rev. D61, 2000 (1998).

\bibitem{gribov69}
V.N. Gribov, Sov. Phys. JETP 29 (1969) 483.


\bibitem{karmanov73}
V.A. Karmanov and L.A. Kondratyuk, JETP Lett. 18 (1973) 266.

\bibitem{pillerRev}
G. Piller and W. Weise, Phys. Rep. 330 (2000) 1; [hep-ph/9908230].

\bibitem{goulianos99}
K. Goulianos and J. Montanha, Phys. Rev. D59 (1999) 114017; [hep-ph/9805496];
K. Goulianos, Phys Lett B358 (1995) 379; Erratum: ib. B363 (1995) 268.  

\bibitem{grassberger77} 
P.G. Grassberger, Nucl. Phys. 125B (1977) 83.

\bibitem{miettinen78}
H.I. Miettinen and J. Pumplin,  Phys. Rev. 18D (1978) 1696.





\end{thebibliography}
\end{document}